\documentstyle[12pt]{article}
\begin{document}

\title{A Bound on Violations of Lorentz Invariance}
\author{R. Cowsik$^1$  \& B.V.Sreekantan$^2$ \\ 
$^1$Indian Institute of Astrophysics, Bangalore 560 034\\
$^2$ National Institute of Advanced Studies, Bangalore 560 012}
\date{}
\maketitle

\baselineskip 20pt
{\it Recently Coleman and Glashow [1] have developed a model which allows the
introduction of a small violation of Lorentz invariance.  Observational
signatures arise because this interaction also violates flavor
conservation and allows the radiative decay of the muon,  $\mu \rightarrow
 e+ \gamma$, whose branching ratio increases as b $\gamma^4$ where
$\gamma$ is the Lorentz factor of the muon with respect to the reference
frame in which the dipole anisotropy of the universal microwave radiation
vanishes.  In this paper we place a bound of b$< 10^{-25}$ based on
observations of horizontal air showers with $n_e \geq 5 \times 10^6$.  With
such small values of b the proposed radiative decay of the muon will not
affect the functioning of the muon collider. (THIS IS A PRELIMINARY
VERSION)}

To test by  experiments the limits of validity of Lorentz
invariance or indeed any of the fundamental principles of physics we need
a theoretical model which assumes a specific form for the violation and
makes predictions of physical phenomena which can be searched for by
the experiments [2-5].  The recent model of Coleman and Glashow incorporates tiny
departures from Lorentz invariance which do not also respect flavor
conservation [1].  One of the signatures of such a flavor non conservation is the
transition $\mu \rightarrow e+ \gamma$ whose rate increases rapidly with the
energy of the muon as measured in a preferred frame such as the one in which the
2.7$^o$K universal microwave background does not have any dipole
anisotropy.  Following their suggestion we calculate the possible
contributions of such a process to the flux of ``horizontal air showers'' and
$\mu$-less showers which provide useful estimates for the possible strength
of such an interaction and also provide a good bound on such violations.

The idea on which the bound on flavor violating interactions is derived 
becomes clear by noting that the primary cosmic rays consist mainly of
nuclei which interact strongly when they are incident on the top of the
earth's atmosphere.  The amount of shielding provided by the atmosphere in the
vertical direction above the earth is about 1000 g cm$^{-2}$ and increases as
the secant of the zenith angle $\theta$ upto $\sim 80^o$.  The total grammage
in the horizontal direction is about 36500 g cm$^{-2}$.  The primary cosmic
rays interact in the atmosphere and create a `nuclear active' cascade.  Since
the atmosphere is tenuous with a scale height h $\approx 7 
\times 10^5$ cm pions and kaons in the cascade decay producing the cosmic-ray
muonic component.  Nuclear interactions of pions and kaons with the atmosphere
compete with their decay and become increasingly dominant as the particle
energy increases, so that the spectrum of the muonic  component at high
energies is steeper
than that of the nuclear active component by a factor E$^{-1}$.  Also the muon
component at high energies increases as $\sim$ sec $\theta$, as the scale
height of the atmosphere also has this dependence.  Since the interaction
mean free path of the hadronic components is \hbox {$\sim$ 70 g cm$^{-2}$,}  after
reaching their maximum development, they  are absorbed with 
an absorption mean free path of $\sim$ 100 g cm$^{-2}$.  In contrast the muons
suffer only electromagnetic interactions and propagate with hardly any
reduction in flux.  Now note that as we move away from the vertical towards
the horizontal direction, with increasing sec $\theta$ the nuclear active
components get severely absorbed but the high energy muonic component
increases as $\sim$ sec $\theta$!  Thus at large angles we have a nearly  pure
beam of high energy muons, traversing distances of the order of few times the
scale height $h_\theta \sim h sec \theta$.  Now should the muon decay
radiatively the decay products e and $\gamma$ will induce an electromagnetic
cascade which can easily be observed signalling the violation of flavor
conservation, as described in the model of Glashow and Coleman.  Indeed as the
energy of the muon increases the observability of the e$\gamma$-cascade
increases as it penetrates deeper, spreads wider and produces more observable
electrons and photons.  The electromagnetic cascade has a very broad peak at
about 500 g cm $^{-2}$  from the point of initiation for an electron or
$\gamma$ of energy E $\sim 10^4 $ GeV and the depth of maximum increases
logarithmically with energy.  The total number of electrons at the peak of an
electromagnetic cascade is approximately equal to the energy of the initiating
electron or gamma ray in GeV units.  Thus any array of particle detectors
deployed to detect extensive air showers will be able to detect such showers
generated by the radiative decay of the muon.  There will be negligible amount
of nuclear active particles and muons in these showers.  The background due to
showers induced by the primary cosmic ray nuclei become negligible as we go to
large zenith angles.  Thus  `$\mu$-less' showers appearing in near
horizontal directions constitute a signal of the new process described by
Coleman and Glashow.

To quantify these ideas we note that the spectrum of muons at high
energies near the earth may be parametrized as

$$\mu (E) =  \frac{\kappa_i sec \theta}{E^{\beta +1}_i} cm^{-2} s^{-1} sr^{-1}
GeV^{-1} \eqno (1)$$
with
$$ \kappa_1 = 10 ,  \beta_1 = 2.7 \,\,\,\, for \,\,\,\,  1000 GeV <E<
10^5\,\,\,   GeV \eqno (2) $$
and
$$\kappa_2 = 10^4, \beta_2 = 3.3 \,\,\,\,\, for \,\,\,\,  10^5 GeV
<E< 3 \times 10^7 GeV \eqno (3)$$

\noindent
Here $\beta_1$ and $\beta_2$ are the power law exponents of the primary
cosmic ray
spectrum at energies of 10 to 30 times the energy of the muon.

According to Coleman and Glashow[1] the total decay probability per unit
time, $\Gamma$, of a muon of Lorentz factor $\gamma$ is given by:

$$\Gamma = \Gamma_w + \Gamma_r = \frac{1 + b \gamma^4}{\gamma \tau_o} =
\frac{1}{\gamma \tau_o} + \frac{b \gamma^3}{\tau_o} \eqno (4)$$

Here $\tau_o \approx 2.2 \times 10^{-6}s$ is the life-time of the muon and
b is a very small parameter describing the violation of Lorentz
invariance and flavor conservation.  For a muon decay close to
the earth say within a distance d of about 5 km ($\sim 700 g cm^{-2}$ from
the air shower array), it has to survive decay during its flight though the
atmosphere beyond this i.e. a distance of few times  $h_\theta$, the scale
height in that direction.  Thus the number of muons decaying in the 5 km
stretch is given by

$$s(E) \approx \kappa sec \theta E^{- \beta -1} exp \left\{ - j h_\theta
\Gamma/c \right\} \Gamma d/c \eqno (5) $$

\noindent
where j is a number of the order of 2 to 3.  Noting that $\Gamma$ is a
small number and that at high energies $\Gamma \sim \Gamma_{r}$, the
exponential in eq. 5 may be set to unity and eq. 5 is rewritten as

$$s(E) \sim \kappa sec \theta E^{- \beta - 1} . \Gamma_r d/c \approx
\frac{\kappa sec \theta b d m^{-3}_\mu}{c \tau_o} E^{2-\beta} \eqno (6) $$

$$ \equiv  \kappa b \eta \,\, E^{2-\beta}$$

\noindent
where $\eta = d m^{-3}_\mu sec\theta/c\tau_o \,\, GeV^{-3} \approx 5 \times 10^4
\,\, GeV^{-3}$.
The products of the radiative decay of the muon generate an extensive air
shower which contains a large number of electrons near the maximum, n,
related to the muon energy through the simple relation

$$ n_e \approx E/\epsilon \eqno (7) $$

\noindent
where $\epsilon \approx $ 1 GeV for an electromagnetic shower of primary
energy in the range $10^4$ GeV - $10^6$ GeV.  The number spectrum of
particles that will be seen by an air shower array is given by

$$f(n) \approx \epsilon^{3-\beta}. \kappa b  \eta  \, n^{2-\beta}$$

Or the number of showers F, of size larger than n is given by

$$ F(n) = \int ^\infty_n \, f(n^\prime)dn^\prime \eqno (9) $$

$$F_2(n) = \frac{\epsilon^{3-\beta} \kappa_2 b \eta}{\beta_2-3}
n^{3-\beta} \,\,\,\, for \, n
\geq 10^5 \eqno(10)  $$

$$F_1(n) = \frac{\epsilon^{3-\beta} \kappa_1 b \eta }{3-\beta_1}
\left[10^{5(3-\beta)} -
n^{3-\beta} \right ] + F_2(10^5) \,\,\,\, for \,\, n < 10^5 \eqno (11)$$

We compare the integral number spectrum of horizontal air showers obtained
by Nagano et al [6] with the Akeno array in Fig. 1 for $b=10^{-23}$ (curve a)
and $b=10^{-25}$ (curve b).  Note that $b\sim 10^{-23}$ excluded even by
the lower energy data at $n_e \sim 10^5$ and the bound

$$b<10^{-25} \eqno (12)$$

\noindent
obtains when we consider the fluxes of horizontal air showers quoted  by
Nagano et al for $n_e \sim 5 \times 10^6$.  Clearly these bounds are
considerably more stringent than those derived by looking at the depth
intensity curves for muons and as such small values of branching ratio for
radiative decay will not have any detrimental effects on the functioning
of muon colliders (Coleman and Glashow 1998).  It is interesting to note
that in the Coleman Glashow model this limit translates to

$$\mid 1-c \mid \leq 6 \times 10^{-21} \eqno (13)$$

This limit is of course several others weaker than those reviewed in their
paper.

\newpage\

{\bf Figure Caption}

Fig. 1 The integral flux of horizontal air showers given by Nagano et al.
is compared with the expectation from the Coleman-Glashow process for the
two values of b, $10^{-23}$ and $10^{-25}$ respectively.  

\begin{thebibliography}{99}


\bibitem{1} S.Coleman and S.L.Glashow, Phys. Lett. B \underbar {405}, 249 (1997);
Harvard University Theoretical Physics Preprint 98/AO76 (prt. comm).

\bibitem{2}  C.M. Will, Theory and Experiment in Gravitational Physics
(Cambridge University Press, Cambridge, 1993)

\bibitem{3} M.I. Haugan and C.M.Will, Physics Today, \underbar {40} (May 1987)

\bibitem{4} E.Fischbach, M.P.Haugan, D.Tadic and H.Y.Chang, Phys. Rev. 
\underbar {D 32} (1985) 154.

\bibitem{5} G.L.Greene, M.S.Dewey, E.G.Kessler, Jr and E.Fischbach, Phys.
Rev. \underbar {D 44} (1991) 2216.

\bibitem{6} M.Nagano, H.Yoshii, T.Hara, N.Hayashida, M.Honda, K.Kamata,
S.Kawaguchi, T.Kifune, Y.Matsubara, G.Tanahashi and M.Teshima, J.Phys. G:
Nucl. Phys. \underbar {12} (1986) 69-84.
\end{thebibliography}
\end{document}